\documentclass[prd,a4paper,preprint,showkeys]{revtex4}
\usepackage{graphicx}

\begin{document}

%My commands
%%this area is reserved to define new-commands.
\newcommand{\be}{\begin{equation}}
\newcommand{\ee}{\end{equation}}
\newcommand{\bq}{\begin{eqnarray}}
\newcommand{\eq}{\end{eqnarray}}
\newcommand{\bsq}{\begin{subequations}}
\newcommand{\esq}{\end{subequations}}
\newcommand{\bc}{\begin{center}}
\newcommand{\ec}{\end{center}}

\title{Cycloops: Dark Matter or a Monopole Problem for Brane Inflation?}

\author{A. Avgoustidis}
\email[Electronic address: ]{A.Avgoustidis@damtp.cam.ac.uk}
\affiliation{Department of Applied Mathematics and Theoretical Physics,
Centre for Mathematical Sciences,\\ University of Cambridge,
Wilberforce Road, Cambridge CB3 0WA, United Kingdom}
\author{E.P.S. Shellard}
\email[Electronic address: ]{E.P.S.Shellard@damtp.cam.ac.uk}
\affiliation{Department of Applied Mathematics and Theoretical Physics,
Centre for Mathematical Sciences,\\ University of Cambridge,
Wilberforce Road, Cambridge CB3 0WA, United Kingdom}

\begin{abstract}
 We consider cosmic loop production by long string interactions in   
 cosmological models with compact extra dimensions. In the case that  
 the compact manifold is not simply connected, we focus on the   
 possibility of loops wrapping around non-trivial cycles and becoming  
 topologically trapped. Such loops, denoted cycloops, behave like matter   
 in the radiation era, posing a potential monopole problem. We calculate    
 the number distribution and the energy density of these objects as   
 functions of cosmic time and use them to study cosmological  
 constraints imposed on simple brane inflation models. For typical   
 choices of parameters we find that to avoid cycloop domination before   
 the matter-radiation transition, the strings must be unacceptably   
 light, namely $G\mu<10^{-18}$, unless some mechanism to dilute the   
 cycloops is provided. By exploring the full parameter space however,    
 we are able to find models with $G\mu \sim 10^{-14}$, which is marginally   
 consistent with brane inflation. In such models the cycloop could   
 provide an interesting dark matter candidate.          
\end{abstract}

\preprint{}
\maketitle

\section{\label{intro}Introduction}
 
 There has been a notable revival of interest in cosmic strings recently   
 both for theoretical and observational reasons (for a review see Ref.   
 \cite{Kibblerev}). The main motivation comes from brane inflation   
 \cite{DvalTye,BMNQRZ}, where cosmic strings are produced at the end of an   
 inflationary phase \cite{SarTye,Vilenk}, and in particular the proposal   
 that these strings can be fundamental (super)strings (F-strings),   
 D-strings, or bound states between them \cite{PolchStab,Vilenk}.  
 The fact that cosmic strings can be produced as topological defects in   
 brane inflation is particularly exciting because such strings could be   
 observable \cite{JoStoTye2,PogTWW,DamVil}, providing a potentially   
 observational handle on superstring physics \cite{PolchRevis, PolchIntro}.  
 Indeed, cosmic strings appearing in these models can be substantially   
 different than usual strings. They can be very light ($10^{-12} < G\mu <   
 10^{-7}$) and can have an intercommuting probability much less than unity   
 \cite{JoStoTye2, PolchProb}. For such strings, interactions are suppressed  
 leading to a much greater string number density today. More complicated   
 `entangled' networks, with strings joining at trilinear vertices, can  
 also be formed \cite{Vilenk, PolchStab}. 

 The evolution of these strings was first discussed in terms of a one-scale  
 model in Ref. \cite{JoStoTye2} (also see Ref. \cite{Vilenk}). More recently,  
 an extension of the Velocity-dependent-One-Scale (VOS) model \cite{vos0,  
 vos,vosk} for theories with extra dimensions has been presented (the EDVOS  
 model) \cite{extvos}, which can be used for more quantitative estimates of  
 scaling string densities. The evolution of non-intercommuting and entangled   
 networks within a VOS model approximation was studied in \cite{nonint} (see 
 also~\cite{MTVOS}). Evolving strings with small intercommuting probability   
 has also been discussed in Refs. \cite{VilSak, DamVil, Sak}.   
      
 The purpose of this paper is to study the evolution of closed cosmic  
 strings (loops) in the presence of extra compact dimensions. As noted  
 in Ref. \cite{Vilenk}, if the compact manifold is not simply connected,  
 such loops can wrap around non-trivial cycles and become topologically   
 trapped. These objects will behave like monopoles and could pose   
 constraints on the models which have them. Similar monopole-like  
 objects have also been discussed recently in Refs. \cite{BarnBCS,  
 MatsNecl}, but their cosmological evolution or the constraints they 
 impose have not been fully investigated. Further constraints exist 
 on a class of cosmic superstring models, for example from dilaton  
 emission \cite{Babichev}.   

 The paper is organised as follows. In section \ref{loops} we briefly   
 review the evolution of cosmic loops in three spatial dimensions and   
 then focus on the case where extra compact dimensions are present. We    
 consider the possibility of loops winding around the extra dimensions   
 and show that their evolution can be very different from usual loops,   
 if the compact manifold admits non-trivial 1-cycles. In particular, 
 their energy density scales like matter in the radiation era, posing   
 a potential monopole problem. In section \ref{cosmconstr} we consider   
 the cosmological constraints imposed on simple brane inflation models   
 by requiring the absence of such a monopole problem. Our conclusions   
 are summarised in section \ref{conc}.

\section{\label{loops}Cosmic Loops}
  
 Numerical simulations of cosmic string formation after symmetry breaking  
 phase transitions suggest that roughly $80\%$ of the total string length   
 is in the form of infinite Brownian strings, while the remaining $20\%$     
 is in closed loops \cite{VachVil}, though these ratios depend somewhat  
 on the topology of symmetry breaking (see \cite{book} and references  
 therein). These loops normally oscillate and radiatively decay, but   
 more loops are dynamically produced by long string interactions. Indeed,   
 a string curling back on itself or two colliding strings can chop off   
 one or more small loops. Simulations of string evolution verify this   
 picture, though the evidence for linear scaling of loop sizes is only   
 now emerging relative to the smallest possible size allowed by the   
 numerical resolution \cite{BenBouch,AllShel,VanOlumVil}. Such loops   
 typically have a very small cross-section and decouple from the long   
 string network, thus taking energy away from it. This mechanism   
 is crucial in ensuring that the network achieves a scaling solution,   
 that is, one in which the correlation length remains constant with   
 respect to the horizon size \cite{Kibble,book}. In the following we   
 will briefly review how the loop energy density evolves in time in   
 the standard picture and then identify the possibility of distinctively   
 different evolution in the presence of compact extra dimensions.

 \subsection{Standard Loops}

  The scaling property of the long string network allows one to describe  
  the production of loops in terms of a scale-invariant function  
  $f(\ell/L)$ \cite{Kibble}. This is defined so that the energy loss into  
  loops of size between $\ell$ and $\ell + \delta\ell$ per correlation   
  volume per unit time is $\mu f(\ell/L) \delta\ell/L$, where $\mu$  
  is the cosmic string tension and $L$ the correlation length, defined  
  so that the energy density of the long string network is given by    
  $\rho_\infty=\mu/L^2$. One then defines a loop number density distribution  
  $n(\ell,t)$ so that $n(\ell,t) \delta\ell$ gives the number density of   
  loops in the length range $\ell$ to $\ell + \delta \ell$. A loop energy   
  density distribution can also be defined by  
   \be\label{rhodef}   
    \rho(\ell,t)=\mu\ell n(\ell,t)\,.  
   \ee 
  Taking into account the dilution due to Hubble expansion, an equation   
  for the rate of change of the loop energy density distribution can be  
  written  
   \be\label{rhodot}  
    \dot\rho(\ell,t)=-3\left(\frac{\dot a}{a}\right) \rho(\ell,t) +  
    g \frac{\mu}{L^4} f(\ell/L)  
   \ee                
  where $a(t)$ is the scalefactor and $g$ a Lorentz factor accounting
  for non-zero center-of-mass kinetic energies of the produced loops.      
  Equation (\ref{rhodot}) can be integrated to give at late times in 
  the radiation era \cite{book} 
   \be\label{rho} 
    \rho(\ell,t)=\frac{\mu g\gamma^{-5/2}}{(t \ell)^{3/2}} 
    \int_0^{\infty} \sqrt{x} f(x) dx\,.      
   \ee
  where $\gamma$ is defined from $L=\gamma t$ and is constant for a   
  scaling network.     
 
  We now use an approximation in which all loops are produced at a constant
  size relative to the horizon i.e. $\ell=\alpha t$. Then the loop production
  function becomes a $\delta$-function
   \be\label{fofx}
    f(x)=\tilde c \delta(x-\alpha/\gamma)
   \ee
  so that
  \be\label{nu_r}
   \nu_r = g \gamma^{-3} \alpha^{1/2} \tilde c \,.
  \ee
  For standard strings the constant $\tilde c$ (the loop production
  parameter) can be extracted from simulations and is of order unity
  \cite{vos}. From equations (\ref{rhodef}), (\ref{rho}) and (\ref{nu_r})
  we can read the loop number density distribution in this approximation
   \be\label{n_li}
    n(\ell,t) = \frac{g \tilde c \gamma^{-3} \alpha^{1/2}}
    {t^{3/2} \ell^{5/2}}\,.
   \ee

  So far we have assumed that loops have constant length $\ell$. We know  
  however that they lose energy (and hence length) due to radiative  
  processes. Assuming that the dominant decay mechanism is emission of   
  gravitational radiation, a loop of initial size $\ell_i$ (formed at  
  time $t_i$) will have a length $\ell = \ell_i - \Gamma G \mu (t-t_i)$  
  at a later time $t$, where $\Gamma$ is a constant of order 65 \cite{book}. 
  Since equation (\ref{n_li}) is expressed in terms of the length at which   
  the loops are produced, we can write for the number density distribution  
  at time $t$  
   \be\label{n_l}  
    n(\ell,t) = \frac{g \tilde c \gamma^{-3} \alpha^{1/2}}  
    {t^{3/2} [\ell+\Gamma G \mu (t-t_i)]^{5/2}} \,.  
   \ee            
  Because of the $\delta$-function approximation (imposing   
  $\ell_i=\alpha t_i$) all loops of a given length $\ell$ at time $t$  
  were produced at the same time $t_i$ (when they had the same initial  
  length $\alpha t_i$). For small loops at late times $t>>t_i$ in   
  (\ref{n_l}) but for large loops (of length comparable to $\alpha t$)    
  $\ell$ is similar to its initial size $\ell_i$ and $t_i$ cannot be    
  neglected. The maximum size of loops at time $t$ is $\alpha t$ and    
  corresponds to the loops that have just been created ($t_i=t$).              

  The dominant contribution to $n(\ell,t)$ is from the smallest loops  
  so that we can ignore $t_i$ in equation (\ref{n_l}). For $\ell>\Gamma   
  G \mu t$ the distribution is dominated by small $\ell$, but for   
  $\ell<\Gamma G \mu t$ it is linear in $\ell$, so that the dominant   
  contribution comes from loops of size $\ell\sim \Gamma G \mu t$.    
  We therefore have, by integrating $n(\ell,t)$ over all loop sizes,   
   \be\label{nstand}  
    n(t)\approx \left. \frac{2}{3} \frac{g \tilde c \gamma^{-3}   
    \alpha^{1/2}}{(t\ell)^{3/2}} \right|_{\ell=\Gamma G\mu t} = \frac{2}{3}  
    \frac{g \tilde c \gamma^{-3} \alpha^{1/2}}  
    {(\Gamma G \mu)^{3/2}\,t^3}\,.        
   \ee  
  The loop energy density at time $t$ in the radiation era is similarly   
   \be\label{rhostand}  
    \rho(t) \approx 2\left(\frac{\alpha}{\Gamma G \mu}\right)^{1/2} \frac{  
    g \tilde c \mu}{\gamma^3 t^2}\, ,  
   \ee    
  which scales like the (radiation) background density. Eventually, matter   
  will dominate and the loops will only contribute a negligible amount  
  to the energy density of the universe.  

  However, in theories with extra compact dimensions it is possible that   
  loops can wrap around 1-cycles in the compact manifold and become   
  topologically trapped, not being able to shrink to zero size \cite{Vilenk}. 
  We call such objects `cycloops' that is, loops winding around non-trivial    
  cycles. One expects the energy density of these objects to scale   
  like matter so the possibility that they can dominate the universe   
  arises. We consider cycloops in more detail below.

 \subsection{\label{cycloops}Cycloops: a monopole problem for brane inflation}

  We now consider the case where the string network evolves in a   
  $(D+1)$-dimensional spacetime with three large FRW spatial dimensions  
  and the rest compactified at a scale $R$, which is less than the    
  correlation length $L$. We can still describe this situation by an    
  effective three dimensional model in which we introduce an  
  intercommuting probability to account for the fact that strings   
  can miss each other, because of the presence of extra dimensions  
  \cite{JoStoTye2, Vilenk, extvos}. The discussion in the previous   
  section follows up to equation (\ref{n_li}) but now the loop     
  production parameter $\tilde c$ is suppressed by a factor   
   \be\label{P}  
    P \sim \left( \frac{\delta}{R} \right)^{D-3}  
   \ee 
  where $\delta$ is the effective string thickness, a capture radius  
  for string interactions (for a calculation of P in string theory see  
  \cite{PolchProb}). 

  The scaling parameter $\gamma$ is also suppressed by a factor of $P$  
  within the one-scale approximation \cite{JoStoTye2} but   
  small-scale-structure on strings is expected to lead to a weaker $P$  
  dependence. This can be incorporated in the one-scale or VOS model  
  by introducing an effective intercommuting probability $P_{\rm eff}=  
  f(P)$ \cite{extvos}. Flat space simulations suggest that $\gamma$   
  is suppressed by a factor of $\sqrt{P}$ rather than $P$ \cite{Vilenk,    
  VilSak, Sak} and simulations in expanding space are in progress   
  \cite{inpreparation}.   
     
  Thus, with the understanding that $\tilde c$ and $\gamma$ are   
  suppressed compared to their standard three-dimensional values,   
  the number density distribution of loops with respect to their  
  initial length is given by (\ref{n_li}). These loops can radiate   
  energy reducing their length, but, unlike the three-dimensional   
  case, not all loops can shrink to zero. If the compact manifold is   
  not simply connected then some loops can wrap around non-trivial   
  1-cycles and become trapped. These loops, the cycloops, can only   
  shrink down to a minimum size equal to their winding number times   
  the size of the non-trivial cycle. They will therefore look like    
  stable monopoles to a three-dimensional observer. We note that 
  similar objects were recently found in Ref. \cite{BarnBCS} as a   
  result of {\rm D}-brane annihilation in the case that the   
  correlation length of the produced brane-defects is smaller   
  than the size of the extra dimensions. Ref. \cite{MatsNecl} also  
  discusses monopole-like objects arising as kink configurations on   
  the worldvolume of long strings, associated to the scalar fields 
  that describe the string position in the extra dimensions.     
    
  Since loops are formed at a length $\alpha t$, larger and larger  
  loops will be produced as time increases. As the strings explore the  
  extra dimensions, we expect the winding number to increase with time,   
  so that the minimum length that a cycloop can reach will depend on   
  the time it was formed. (Of course, depending on energetic  
  considerations, a large cycloop of winding number $N$ can break up  
  into $N$ cycloops of single winding.)   

  \subsubsection{\label{winding}Winding Number}    
  We can study the dependence of the winding number on time using   
  the Extra-Dimensional VOS model introduced in \cite{extvos}. In   
  that model there is a parameter $w_\ell$, which provides a measure   
  of the amount of string length in the extra compact dimensions. In   
  particular $w_\ell^2$ is the average value of the ratio of the square    
  of the string length in the $D-3$ compact dimensions, over the square    
  of the length in all dimensions. For a loop of total length $\ell$,    
  the parameter $w_\ell$ can be used to estimate its length in the    
  compact dimensions and this can be turned into a winding number, subject    
  to assumptions about the structure of the string in these dimensions    
  and the non-trivial cycles in the compact manifold.  
  
\begin{figure}
   \includegraphics[width=2.5in,keepaspectratio]{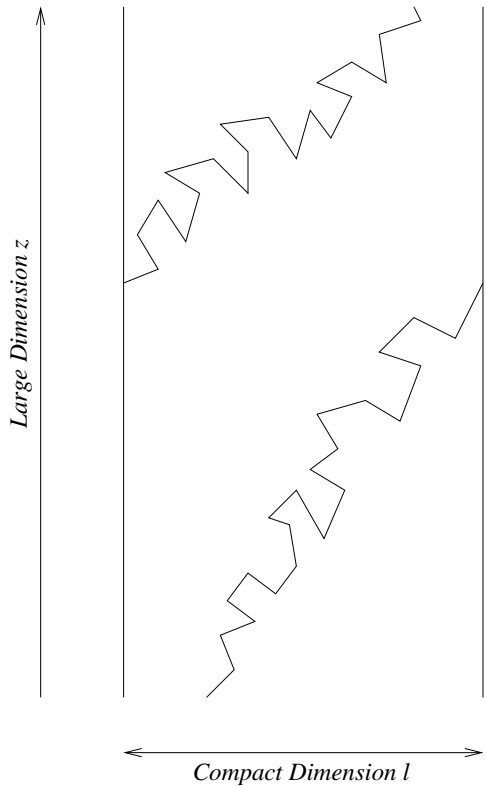}
   \caption{\label{sssfig} Random walk regime: The spatial 
   structure of strings in the extra dimension (assumed Brownian)   
   can give rise to non-trivial windings.}
  \end{figure}

\begin{figure}
   \includegraphics[width=5in,keepaspectratio]{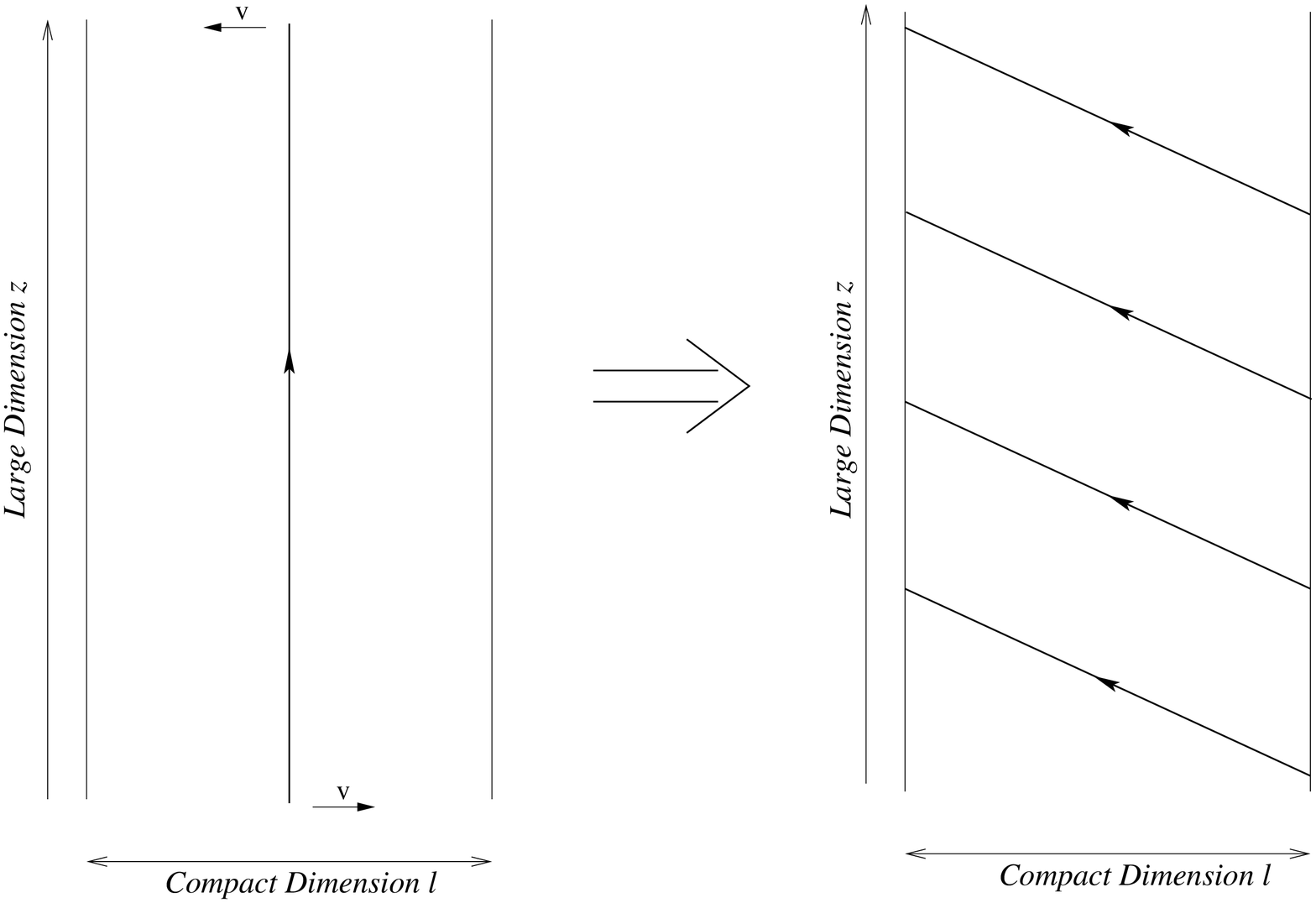}
   \caption{\label{velcorfig} Velocity correlations regime: String   
            velocities in the extra dimensions cannot be correlated    
            at distances greater than the correlation length at  
            formation $L_0$. Thus the velocities at the endpoinds of a   
            string segment of length $L_0$ can be expected to have   
            different directions. This would tilt the string as shown    
            to produce windings.}   
  \end{figure}

  We can identify two distinct scenarios: in the first one, the strings   
  are assumed to form with an initial correlation length $L_0$ smaller 
  than the size of the compact dimensions $R$. We suppose they form a   
  random walk shaped-structure in these dimensions, which can give   
  rise to non-trivial windings (see Fig. \ref{sssfig}). Having $L_0<R$   
  does not appear to be the usual brane inflation scenario (but see Ref. 
  \cite{BarnBCS}); presumably, the Brownian strings slowly migrate off   
  the branes because of transverse momenta arising from the brane   
  collision. On the other hand, for $L_0>R$, we have the scenario in   
  which straighter strings acquire windings more directly because of    
  their transverse velocities. Since these velocities can only be    
  correlated at length-scales less than the correlation length of   
  the network, the strings begin winding around the compact dimensions,  
  as illustrated in Fig. \ref{velcorfig}. We will consider each scenario  
  separately.    

  \noindent {\it (i) Random walk:} In this regime, we assume that     
  string winding arises from the strings' spatial structure in   
  the compact dimensions, which can be modelled as a random walk of  
  step $\xi_\ell$. For a loop of total length $\ell \gg \xi_\ell$,   
  its length in the compact dimensions is approximately $\ell {w_\ell}$.   
  Because of the random walk shape of the string, the total displacement  
  in these dimensions is considerably less, namely $\sqrt{\ell w_\ell   
  \xi_\ell} = \sqrt{\alpha w_\ell \xi_\ell t}$. If this displacement   
  is greater than the size of 1-cycles present in the compact manifold,  
  R, a non-trivial winding can be developed with a winding number given by   
   \be\label{Nfloor}   
     N \simeq \frac{\sqrt{\alpha w_\ell \xi_\ell t}}{R}\,.    
   \ee  
  For a cycloop created at $t=t_i$, the minimum length to which it can shrink   
  is then   
   \be\label{lmin}  
    \ell_{\rm min} \simeq NR \simeq \sqrt{\alpha w_\ell \xi_\ell t}   
    \equiv \kappa \, t_i^{1/2}         
   \ee   
  where we have set $\kappa \equiv \sqrt{\alpha w_\ell \xi_\ell}$. Thus,   
  in this regime, the minimum cycloop length is proportional to the    
  square root of the time at which the loop was formed.    

  \noindent {\it (ii) Velocity correlations:} We now consider a situation  
  where the strings are produced with significant velocity components in   
  the compact dimension \cite{extvos}. These cannot be correlated beyond  
  distances of order the correlation length at string formation $L_0$,    
  and so points on the string separated by distances greater than $L_0$   
  can be expected  to move in different directions. Such motion in 
  the compact dimensions can result in string winding, if the initial   
  correlation length is greater than the size of the non-trivial cycles   
  present in the compact manifold.  

  To quantify this, we estimate that the number of windings per initial   
  correlation length will be given by  
   \be\label{windpercor}  
    N_{L_0}=w_\ell \, \frac{L_0}{R}, \;\;  L_0>R \,.  
   \ee      
  A loop of length $\ell$ will then have an associated winding number  
   \be\label{windloop}  
    N \simeq \left\{
     \begin{array}{ll}
      \sqrt{\frac{\ell}{L_0}} N_{L_0} = w_\ell \frac{\sqrt{\ell L_0}}  
                                       {R} \;\; , \;\; \ell>L_0  \\
      \;\frac{\ell}{L_0}N_{L_0} = w_\ell \frac{\ell}{R}\;\;\;\;\;\;\;\;   
                                        , \;\; \ell<L_0 \,.   
     \end{array}
    \right.  
   \ee 
  Loops are typically produced at a fraction of the correlation length,  
  so the first loops will have a size $\ell<L_0$. Since the correlation    
  length increases with time, loops produced at later times will have a   
  size greater than the initial correlation length at formation. However,  
  the loop number density is, as in the three-dimensional case, dominated    
  by the smallest loops in the distribution (see next section) and so the   
  vast majority of loops will have $\ell<L_0$. Writing $\ell=\alpha t$    
  we obtain for the winding number  
   \be\label{windno}    
    N \simeq \frac{w_\ell \alpha}{R} t  
   \ee   
  so that a cycloop created at time $t_i$ will reach a minimum length  
   \be\label{lminlin}   
     \ell_{\rm min} \simeq NR \simeq w_\ell \alpha t_i \,.  
   \ee   
  This is linear in loop formation time, in contrast to the previous  
  (random walk) regime where $\ell_{\rm min}\!\propto\!t_i^{1/2}$. 
  
  It is convenient to introduce a function $f$ such that  
   \be\label{f}  
    \ell_{\rm min}=f(t_i)= \left\{  
     \begin{array}{ll}  
      \kappa t_i^{1/2}\,\,,\;{\rm random\;walk\;regime} \\    
      w_\ell \alpha t_i\,,\;\!{\rm velocity\;correlations\;regime}   
     \end{array}
    \right.
   \ee 
  Next we study the cycloop number and energy density distribution in both   
  regimes.    
  
  \subsubsection{\label{distribution}Number and Energy Density Distribution}
  We first express the number distribution (\ref{n_li}) in terms of  
  the cycloop length at time $t$ (equation (\ref{n_li}) is expressed in  
  terms of the initial length at production). We recall from last section    
  that the smallest possible cycloops have size $\ell_{\rm min}(t_c)=f(t_c)$   
  where $t_c$ is the time at which the first cycloops form and the function  
  $f$ is either proportional to a square root or linear, depending on whether   
  the winding is produced by random walk spatial structure or velocity    
  correlations. Also, the largest possible loops at time $t$ are the ones  
  just produced with length $\alpha t$. The cycloop distribution is    
  therefore zero outside this range.  
  
  At any time $t$ cycloops fall into two classes: those that have reached   
  their minimum size and those that did not have enough time to do so and   
  are still shrinking. At early times all loops in the distribution are  
  shrinking but at some point $t_{cf}>t_c$ the first loops of initial size   
  $\alpha t_c$ reach their minimum length $\ell_{\rm min}(t_c)$ and freeze.  
  For $t>t_{cf}$ there are both shrinking and stabilised cycloops in   
  the distribution. Thus, for $t>t_{cf}$ the cycloop number density   
  distribution is 
   \be\label{ncycl_l}  
    n_{\rm cycl}(\ell,t) = \left\{  
     \begin{array}{lll}   
      \frac{g \tilde c \gamma^{-3}\alpha^{1/2}}{t^{3/2}   
      [\ell+\Gamma G \mu (t_{if}-t_i)]^{5/2}} & , & \;\;\; \ell_{\rm min}  
      (t_c) < \ell < \ell_{\rm min}(t_m)     \\  
      \;\frac{g \tilde c \gamma^{-3}\alpha^{1/2}}{t^{3/2}   
      [\ell+\Gamma G \mu (t-t_i)]^{5/2}} & , & \;\;\; \ell_{\rm min}(t_m)   
      < \ell < \alpha t    \\ 
      \;\;\;\;\;\;\;\;\;\;\;\;\; 0 & , & \;\;\; \ell < \ell_{\rm min}(t_c)  
      \; {\rm or} \; \ell > \alpha t   
     \end{array}   
    \right.    
   \ee
  where $t_{if}$ is the time at which a loop of initial size $\ell_i=  
  \alpha t_i$ reaches its minimum size $\ell=\ell_{\rm min}(t_i)$ and   
  freezes. This is uniquely determined by $\ell_i=\ell+\Gamma G\mu   
  (t_{if}-t_i)$ giving 
   \be\label{t_if}
    t_{if}=\frac{(\alpha+\Gamma G\mu)t_i-\ell_{\rm min}(t_i)}{\Gamma G\mu}\,.  
   \ee
  Also the time $t_m$, which determines the length where the two   
  distributions in (\ref{ncycl_l}) match, is the time of formation of   
  cycloops that have just reached their minimum size $\ell_{\rm min}(t_m)$   
  at time $t$. This is found by saturating $t_{if}\le t$, which has only   
  one real solution for both functions $\ell_{\rm min}(t_i)$ discussed   
  above.  
  
  The distribution (\ref{ncycl_l}) is dominated by cycloops of smallest  
  initial length, which have already reached their minimum size. Thus  
  it suffices to consider only the frozen cycloops (ignoring all other  
  loops in the distribution), whose number density distribution is given   
  by the first part of equation (\ref{ncycl_l}). Noting that the quantity   
  in the square bracket is equal to $\ell_i=\alpha t_i$ and using equation  
  (\ref{f}) this becomes  
   \be\label{nfrozenback}
    n(\ell,t)=\frac{g \tilde c \gamma^{-3}}{\alpha^2 [f^{-1}(\ell)]^{5/2}   
    t^{3/2}}  
   \ee
  with $f^{-1}$ the inverse function of $f$, equal to $t_i$ by equation   
  (\ref{f}). We now consider the two winding regimes discussed in last  
  section.

  \noindent {\it (i) Random walk:} In this case $f^{-1}(\ell)  
   =(\ell/\kappa)^2$ and equation (\ref{nfrozenback}) becomes  
    \be\label{nfrozensss} 
     n(\ell,t)=\frac{g \tilde c \gamma^{-3} \kappa^5}{\alpha^2 \ell^5
     t^{3/2}}  \,.  
    \ee
  Integrating this over all loop sizes yields the number density of 
  cycloops at time $t$  
   \bq\label{nsss}
    n(t)=\int_{\ell_{\rm min}(t_c)}^{\ell_{\rm min}(t_m)}n(\ell,t)   
    d\ell \simeq \frac{1}{4}\frac{g \tilde c \gamma^{-3} \kappa^5}  
    {\alpha^2 [\ell_{\rm min}(t_c)]^4 t^{3/2}} \nonumber \\  
    =\frac{1}{4} g \tilde c \gamma^{-3} \alpha^{-3/2} w_\ell^{1/2}   
    \frac{\xi_\ell^{1/2}}{t_c^2 t^{3/2}}   
   \eq   
   where we have used the fact that the distribution is dominated by  
   loops of the smallest size so that the integral can be approximated  
   by its lower limit.   

   The cycloop energy density at time $t$ can be found in a similar way   
   \bq\label{rhosss}  
    \rho_{\rm cycl}(t)=\int_{\ell_{\rm min}(t_c)}^{\ell_{\rm min}(t_m)}  
    n(\ell,t) \mu\ell d\ell \simeq \frac{1}{3}\frac{g \mu \tilde c  
    \gamma^{-3} \kappa^5} {\alpha^2 [\ell_{\rm min}(t_c)]^3 t^{3/2}} 
    \nonumber \\  
    =\tilde c \gamma^{-3} \alpha^{-1} w_{\ell} \frac{\xi_\ell \, \mu}   
    {(t_c t)^{3/2}}\,.  
   \eq 
  Note that this depends on $\xi_\ell$, the step of the random walk which   
  models the spatial structure of loops in the extra dimensions.   
                 
  \noindent {\it (ii) Velocity correlations:} In this regime we have 
  $f^{-1}(\ell)=\ell/(w_\ell \alpha)$. The number density distribution  
  (\ref{nfrozenback}) now reads  
   \be\label{nfrozenvcor}
    n(\ell,t)=\frac{g \tilde c \gamma^{-3} \alpha^{1/2} w_\ell^{5/2}}  
    {\ell^{5/2}t^{3/2}}
   \ee 
  The corresponding cycloop number and energy densities at time $t$  
  are now  
   \be\label{nvcor}
    n(t) \simeq \frac{2}{3}\frac{g \tilde c \gamma^{-3} \alpha^{1/2}  
    w_\ell^{5/2}} {[\ell_{\rm min}(t_c)]^{3/2} t^{3/2}} = \frac{2}{3} 
    g \tilde c \gamma^{-3} \alpha^{-1} w_\ell \frac{1}{(t_c t)^{3/2}}  
   \ee
  and  
  \be\label{rhovcor}
   \rho_{\rm cycl}(t) \simeq 2\frac{g \mu \tilde c \gamma^{-3}   
   \alpha^{1/2} w_\ell^{5/2}} {[\ell_{\rm min}(t_c)]^{1/2} t^{3/2}}  
   = 2\tilde c \gamma^{-3} w_{\ell}^2 \frac{\mu}{t_c^{1/2} t^{3/2}}\,.  
  \ee 
  In this case the cycloop energy density does not depend on the  
  parameter $\alpha$. Since $\alpha$ is proportional to $\gamma$ this  
  leads to a different dependence of the cycloop energy density on the  
  effective intercommuting probability $P_{\rm eff}$. The dependence   
  on $w_\ell$ and $t_c$ is also different. These differences will be   
  of significant importance in the next section.        
 
  In both regimes the energy density of cycloops scales like matter   
  in the radiation era. This immediately poses a potential monopole   
  problem (or in even closer analogy a vorton problem \cite{vortform,  
  vortlim}) and requiring consistency with standard cosmology would  
  constrain models that allow cycloops. We consider such constraints   
  in the following section.

\section{\label{cosmconstr}Cosmological Constraints}  

 We have seen that the winding number of cycloops increases   
 monotonically with time, so loops produced at later times will   
 have more windings and hence more length. However, most loops    
 are produced at earlier times, when the density of the long string  
 network is higher, and so the cycloop energy density is dominated   
 by cycloops of the smallest possible winding   
 (\ref{nfrozensss}-\ref{rhovcor}). In the usual scenario where the   
 correlation length at the time of cosmic string formation is much   
 greater than the size of the extra dimensions ($L_0 \gg R$), even   
 the smallest cycloops can have a large winding number.   
 However, if the correlation length at string formation is less   
 than the compactification scale (as in Ref. \cite{BarnBCS}), the first   
 closed loops formed by long string interactions will have no windings   
 and will eventually decay. The cycloop energy density will be    
 dominated by loops with winding number of order one, produced at   
 later times, when the spatial structure will have developed enough   
 to allow non-trivial windings.  

 To quantify the cosmological constraints imposed by cycloops we consider  
 a simple ${\rm D}5-\overline{{\rm D}5}$ inflation model in which two of   
 the dimensions parallel to the branes are compactified on a torus of size 
 $R$. The branes collide and annihilate, producing ${\rm D}3$ and   
 $\overline{{\rm D}3}$ branes, which, if they wrap the same dimensions   
 as the mother branes, can be seen as cosmic strings.     

 We consider the `usual' case where the correlation length is of order  
 $L_0\sim H^{-1}\simeq M_{pl}/M_s^2$, with the string scale $M_s$ set    
 to a low GUT scale by the CMB data \cite{SarTye, PogTWW} and the   
 compactification size $R \sim 10\,M_s^{-1}$. Since the correlation   
 length is much greater than $R$, the Kibble mechanism only takes place  
 in the large three dimensions, so the 3-branes produced must wrap the  
 same compact volume as their mother branes \cite{SarTye}. We can   
 therefore consider them as strings and ignore their internal compact  
 dimensions. However, there are also compact dimensions transverse to  
 the branes, in the bulk where the strings can move. Long string    
 interactions can therefore produce cycloops wrapping around these    
 dimensions.    

 We consider the two winding regimes of section \ref{winding} separately:  
  
 \noindent {\it (i) Random walk:} Here the cycloop winding number is    
 determined by (random walk shaped) spatial structure in the extra  
 dimensions, and the cycloop energy density as a function of formation    
 time is given by equation (\ref{rhosss}). To account for changes in    
 the effective number of degrees of freedom near mass thresholds during  
 the evolution of the early universe, we make use of the entropy density   
 of relativistic species  
  \be\label{entropy}
   s(T)=\frac{2\pi^2}{45} \mathcal{N}_s(T) T^3  
  \ee   
 where $\mathcal{N}_s(T)$ is the effective number of relativistic degrees   
 of freedom for entropy as a function of the background plasma temperature T.   
 
 Neglecting annihilations between cycloops of opposite winding, the ratio   
 of the number density of cycloops to the total entropy density is conserved,  
 so that we can write for the cycloop energy density  
  \be\label{ns}   
   \rho_{\rm cyc}(T)=\mu \int \! n(\ell,t_f)\ell d\ell \frac{s(T)}{s(T_f)}  
    =\mu \int \! n(\ell,t_f)\ell d\ell \frac{\mathcal{N}_s(T)}  
     {\mathcal{N}_s(T_f)} \left(\frac{T}{T_f}\right)^3   
  \ee  
 where $t_f$ (resp. $T_f$) denotes the time (resp. temperature) at which   
 the first cycloops, produced at time $t_c$, freeze to their minimum size.   
 Note that we assumed no significant change in the effective number of   
 degrees of freedom between times $t_c$ and $t_f$.    
 
 Writing the energy density of relativistic species as   
  \be\label{rhorel}  
   \rho_{\rm rad}(t)=\frac{\pi^2}{30} \mathcal{N}(T) T^4  
  \ee  
 where $\mathcal{N}(T)$ is the effective number of relativistic degrees
 of freedom for energy density, we form the ratio of $\rho_{\rm cyc}$    
 to the critical density of the universe $\rho_{\rm crit}$ at   
 matter-radiation equality, finding   
  \be\label{rhoccsss}  
   \frac{\rho_{\rm cyc}(T_{\rm eq})}{\rho_{\rm crit}(T_{\rm eq})} =  
   \frac{1.48}{\pi^2} g {\tilde c} \gamma^{-3} \alpha^{-1} w_\ell   
   \,\mathcal{N}_s(T_s)^{1/2}\, \frac{\mathcal{N}_s(T_{\rm eq})}  
   {\mathcal{N}(T_{\rm eq})} \left(\frac{\mathcal{N}_s(T_s)}  
   {\mathcal{N}_s(T_f)}\right)^{1/4} \left(\frac{\xi_\ell \, \mu   
   T_s^3}{T_{\rm eq} M_{pl}^3} \right)   
  \ee  
 Here we have used equation (\ref{rhosss}) and assumed instantaneous   
 reheating to a radiation dominated universe immediately after the end   
 of inflation, so that cosmic time can be related to the plasma temperature   
 $T$ by   
  \be\label{tofT}   
   t=1.51 \mathcal{N}_s(T)^{-1/2}\frac{M_{pl}}{T^2}   
  \ee
 where $M_{pl}=2.44 \times 10^{18}$~GeV is the reduced Planck mass, related 
 to the Planck mass $m_{pl}=1.22 \times 10^{19}$~GeV by $m_{pl}=\sqrt{8\pi}  
 M_{pl}$. We have also assumed that the first cycloops are formed at the   
 time of cosmic string formation $t_s$ (just after brane collision and   
 the end of inflation) so that we can write $T_c\simeq T_s$, which is   
 approximately equal to the reheating temperature.    

 A constraint on the model can be imposed by requiring that cycloops do    
 not dominate the energy density of the universe before the time of equal  
 matter-radiation $t=t_{\rm eq}$ that is,   
 demanding     
  \be\label{demand}  
   \frac{\rho_{\rm cyc}(T_{\rm eq})}{\rho_{\rm crit}(T_{\rm eq})} \le 1 
  \ee 
 For typical models the string tension $\mu$ is of order the string   
 scale $M_s$ squared (see for example \cite{Vilenk}) and assuming   
 instantaneous reheating we have $M_s \sim T_s$. In the random walk  
 scenario, the stepsize $\xi_\ell$ in the compact dimension   
 cannot be smaller than the   
 inverse string scale or larger than the size of the compact dimension  
 $R \sim 10 M_s^{-1}$. Thus we have to assume $\xi_\ell \sim M_s^{-1}   
 \sim T_s^{-1}$. We can also take $\mathcal{N}_s(T_s)$ of order $10^3$,
 as in some GUT models. We will take the effective intercommuting    
 probability to lie in the range $10^{-3} < P_{\rm eff} < 1$. (Note   
 that strictly speaking it is the actual probability $P$ which spans     
 this range \cite{PolchProb}, so the corresponding range for    
 $P_{\rm eff}$ depends on the functional form $P_{\rm eff}=f(P)$,    
 which is still uncertain.)        

 Usual field theory strings have typically $\tilde c \sim 1$ but, as   
 discussed above, in our setup $\tilde c$ is of order the effective  
 intercommuting probability $P_{\rm eff}$. The extra-dimensional VOS   
 model \cite{extvos} suggests $\gamma \sim \tilde c \sim P_{\rm eff}$.   
 Numerical simulations of string networks provide evidence that loop  
 production is typically peaked at sizes of order $10^{-2}-10^{-3}$ 
 the correlation length \cite{BenBouch,AllShel}. Thus we will assume  
 $\alpha\sim 10^{-2}\gamma \sim 10^{-2}P_{\rm eff}$. Note however,   
 that this assumption is only valid once the long string network  
 has reached a scaling regime. If cycloops are produced early enough,  
 it may well be that the network has not yet achieved scaling, in  
 which case $\alpha$ can be considerably larger (we will return to   
 this point later). From numerical simulations, the Lorentz factor  
 $g$ can be taken to be $g\simeq 1/\sqrt{2}$ \cite{book}.   

 Finally, we need to choose some value for the parameter   
 $w_\ell$, which, as mentioned above, is a measure of the   
 relative amount of string length in the extra dimensions.  
 Unfortunately, this is somewhat uncertain and as yet not fully   
 understood since this parameter is determined by the nature of  
 the brane collision. For example if the collision is violent,   
 one expects the strings formed to have significant velocities in   
 the extra dimensions, which would force them to explore these   
 dimensions. On the other hand, a perfectly adiabatic collision  
 would give rise to smooth strings lying on the plane of brane  
 collision. We will take $w_\ell \sim 0.1$, corresponding to  
 strings with approximately one tenth of their length in the extra  
 dimensions.     

 With this choice of parameters and for $P_{\rm eff}\simeq 10^{-2}$   
 the bound (\ref{demand}) becomes  
  \be\label{constrsss}  
   10^8 \frac{T_s^4}{T_{\rm eq} M_{pl}^3} \le 1~\Longrightarrow~  
   T_s \le (10^{-8} T_{\rm eq} M_{pl}^3)^{1/4} \sim 10^{9}~{\rm GeV}.   
  \ee  
 This is orders of magnitude lower than the typical energy scale of these  
 unwarped brane inflation models (GUT scale) and thus such models would be   
 ruled out, unless some other mechanism (e.g. a short period of subsequent   
 inflation \cite{BurEMMM}) operates to reduce the cycloop energy density.  
 
 \noindent {\it (ii) Velocity correlations:} In this case the energy  
 density of cycloops is given by equation (\ref{rhovcor}). Working  
 as before we find 
  \be\label{rhoccvcor}   
   \frac{\rho_{\rm cyc}(T_{\rm eq})}{\rho_{\rm crit}(T_{\rm eq})} =
   \frac{13.3}{\pi^2} g {\tilde c} \gamma^{-3} w_\ell^2   
   \left(\frac{\mathcal{N}_s(T_s)}{\mathcal{N}_s(T_f)}\right)^{1/4}  
   \frac{\mathcal{N}_s(T_{\rm eq})}{\mathcal{N}(T_{\rm eq})} \left(
   \frac{\mu T_s}{T_{\rm eq} M_{pl}^2} \right)
  \ee     
 leading to the constraint (again with typical choice   
 of parameters)  
  \be\label{constrvcor}  
   10^2 \frac{T_s^3}{T_{\rm eq} M_{pl}^2} \le 1 ~\Longrightarrow~  
   T_s \le (10^{-2} T_{\rm eq} M_{pl}^2)^{1/3} \sim 10^{8.5}~{\rm GeV}.  
  \ee    
  
 It is interesting to comment on the differences between the constraints  
 (\ref{constrsss}) and (\ref{constrvcor}). In the random walk regime    
 (\ref{constrsss}) the temperature which saturates the bound scales  
 like $(T_{\rm eq} M_{pl}^3)^{1/4}$, resembling the usual magnetic monopole   
 constraint. On the other hand, the corresponding scaling in the velocity   
 correlation regime is $(T_{\rm eq} M_{pl}^2)^{1/3}$, which by itself  
 would produce a stronger bound by approximately two orders of  
 magnitude. However, some of this difference is compensated by the   
 different numerical coefficients in (\ref{constrsss}) and   
 (\ref{constrvcor}).  

 Indeed, comparing equations (\ref{rhoccsss}) and (\ref{rhoccvcor}) we  
 observe that in the velocity correlation regime there is no dependence  
 on $\alpha$. Also, $w_\ell$ appears quadratically rather than linearly,   
 while the dependence on $\mathcal{N}_s(T_s)$ is essentially lost. Note  
 that because of the quartic (resp. cubic) root in equation (\ref{constrsss})  
 (resp. (\ref{constrvcor})) the cycloop constraint is not strongly    
 sensitive on the choice of parameters. However, different choices  
 of $P_{\rm eff}$ in the range $10^{-3}< P_{\rm eff}<1$ and $w_\ell$ in   
 the range $10^{-2}<w_\ell<1$ can lead to changes in $(T_s)_{\rm min}$   
 of up to a factor of $10^2-10^3$.   

 Thus in most of the parameter space, the cycloop constraint rules out   
 brane inflation models with a low GUT energy scale or higher that   
 produce stable cosmic strings and allow non-trivial cycles in their    
 compact manifolds. However, there is a small region in the parameter   
 space, corresponding to $P_{\rm eff} \simeq 1$, $\alpha \sim \gamma$    
 and $w_\ell < 10^{-2}$ (in the random walk regime), which is still    
 consistent with an energy scale of order $10^{12}$ GeV. One could    
 imagine such a model arising from a quasi-adiabatic brane collision    
 (thus justifying a small value of $w_\ell$) in which the cycloops are  
 produced before scaling has been reached (hence allowing a large value  
 of $\alpha$). In such a model the cycloop could play the role of dark  
 matter. 

 Finally, we consider the recently proposed scenario \cite{BarnBCS} that   
 the correlation length at string formation is less than the size of the   
 extra dimensions. In that case the 3-branes produced do not have to wrap   
 the compact dimensions so domain wall-like objects can be formed. One   
 expects that these objects would interact to produce strings and indeed   
 toy model simulations in Ref. \cite{BarnBCS} support this. It is also    
 noted that monopole-like objects, namely ${\rm D}1$-branes wrapping the  
 compact dimensions can be formed as well.       
     
 It is clear that the cycloops we considered above, i.e. the closed loops  
 (winding around compact dimensions) produced by long string network   
 interactions, cannot be created soon after the end of inflation in the 
 case of $L_0\ll R$, simply because the correlation length is far too small   
 to allow their formation. The first cycloops will form when the correlation   
 length becomes much greater than the size of the extra dimensions or more   
 specifically, in the random walk regime, when the displacement of   
 loops in these dimensions $\kappa t^{1/2}$ becomes greater than the  
 compactification scale $R$ (note that velocity correlations for $R>L_0$  
 can only give rise to winding in a statistical sense, in which case we 
 recover the random walk regime). Thus, the considerations presented above
 would apply for cycloops produced at a time $t_c$, much later than the 
 end of inflation, when the string network density will be smaller. This
 leads to a smaller cycloop number density albeit with a larger average  
 winding number. The net result is that our considerations constrain 
 (mostly) $T_c$ (i.e. $T$ at cycloop formation) rather than $T_s$ ($T$ 
 at string formation and reheating), so the constraint on $T_s$ itself  
 is weaker. Again, this opens up the possibility that the cycloop can provide  
 a dark matter candidate. Also note that in a realistic model, the scalar   
 fields corresponding to the string position in the extra dimensions should   
 be stabilised at late times \cite{PolchProb}, so cycloop production may  
 not be possible if this localisation takes place at $t<t_c$.

\section{\label{conc}Conclusions}   
  
 We have considered cosmic string evolution in the presence of compact  
 extra dimensions and focused on the possibility that long string  
 interactions can produce closed loops wrapping around compact dimensions.  
 In the case that the compact manifold admits non-trivial 1-cycles, these  
 loops can become topologically trapped and behave like stable monopoles  
 or, in closer analogy, vortons. 

 We have studied the dependence of the winding number of such a loop on   
 the time $t$ when the loop was formed and identified two distinct   
 regimes: one in which windings arise as a result of (random walk shaped)   
 spatial structure in the compact dimensions, and one where they   
 are created by string velocity correlations in these dimensions. In   
 the former case the winding number is proportional to $t^{1/2}$, while  
 in the latter to $t$. In both cases, we studied the evolution of these  
 objects (dubbed cycloops) and we have shown that their energy density   
 scales like matter in the radiation era, the distribution being dominated   
 by cycloops of the smallest possible winding number, produced at the   
 earliest possible times. This immediately gives rise to a potential   
 monopole problem, which can be used to impose constraints on models   
 that allow cycloops.   
 
 Considering these constraints in the case of simple (unwarped   
 compactification) brane inflation models and assuming instantaneous  
 reheating immediately after brane collision, we find that in the   
 majority of the parameter space the energy scale of inflation is   
 required to be less than $10^{10}~{\rm GeV}$. Since the typical scale of  
 these models is a low GUT scale, this constraint rules out these models  
 unless some mechanism (e.g. a secondary inflationary phase \cite{BurEMMM})
 is provided to dilute the cycloop density. There is however a small  
 region in the parameter space, corresponding to an intercommuting   
 probability of order unity and large loops (such large loops could be  
 produced at very early times when the loop size is determined by the   
 initial conditions rather than the asymptotic scaling behaviour of  
 the network), which may still be consistent with low GUT scale brane   
 inflation. In such a model the cycloop may be an attractive dark   
 matter candidate. Another situation in which the cycloops are allowed 
 as dark matter is if they are formed long after the brane collision, 
 in which case the cycloop constraint is alleviated.     

 We stress that our analysis directly constrains only models which allow   
 non-trivial 1-cycles in their compact manifolds (e.g. toroidal   
 compactification), but not, for example, the simplest warped-compactification  
 models with simply-connected throats (see however Ref. \cite{Bl-Pil}, where  
 stable monopole solutions were found in the conifold geometry). The   
 outstanding uncertainty in these estimates remains the dynamics of the   
 brane collision and ensuing string-forming transition.

\begin{acknowledgments}

We are grateful to Carlos Martins for many discussions.
A.A. is funded by EPSRC, the Cambridge European Trust and the Cambridge  
Newton Trust. E.P.S.S. is supported by PPARC, grant number PP/C501676/1.   

As this work was completed, a replacement version of hep-ph/0412290
was put on the archive. This new version appears to contain some   
overlapping results for our relic loop density, though that discussion  
seems to depend on localising potentials in the extra dimensions.

\end{acknowledgments}

\bibliography{Cycloops}

\end{document}